

\tolerance=2000
\hbadness=2000
\overfullrule=0pt
\magnification=1200
\baselineskip=18pt


\font\titlesf=cmssbx10 scaled \magstep2

\font\subchaptsf=cmssbx10 scaled \magstephalf
\font\authorsf=cmss10 scaled \magstep1
\font\sf=cmss10
\font\pl=cmssq8 scaled \magstep1

\def\a{\alpha}
\def\b{\beta}

\def\const{{\rm const\,}}
\def\cosh{{\rm cosh\,}}
\def\coth{{\rm coth\,}}
\def\d{\delta}
\def\Det{{\rm Det\,}}
\def\det{{\rm det\,}}

\def\F{{\cal F}}
\def\frac#1#2{{\textstyle {#1\over #2}}}
\def\g{\gamma}

\def\l{\lambda}
\def\m{\mu}

\def\n{\nu}
\def\na{\nabla}
\def\om{\omega}
\def\P{{\cal P}}

\def\R{{\cal R}}

\def\RR{\rm I\!R}

\def\sinh{{\rm sinh\,}}

\def\tanh{{\rm tanh\,}}
\def\tr{{\rm tr\,}}
\def\Tr{{\rm Tr\,}}
\def\log{{\rm log\,}}
\def\vol{{\rm vol\,}}

\def\sq{\mathchoice{\square{6pt}}{\square{5pt}}{\square{4pt}}
    {\square{3pt}}}
\def\square#1{\mathop{\mkern0.5\thinmuskip\vbox{\hrule\hbox{\vrule
    \hskip#1 \vrule height#1 width 0pt \vrule}\hrule}\mkern0.5\thinmuskip}}

\def\ltextindent#1{\hbox to \hangindent{#1\hss}\ignorespaces}

\def\today{\ifcase\month\or January\or February\or March\or April\or May
    \or June\or July\or August\or September\or October \or November
    \or December \fi\space\number\day, \number\year}
\def\heute{\number\day. {\ifcase\month\or Januar\or Februar\or M\"arz
    \or April\or Mai\or Juni\or Juli\or August\or September\or Oktober
    \or November\or Dezember\fi} \number\year}

\def\rightheadline{\it\title\qquad\hfill\rm\folio}
\def\leftheadline{\rm\folio\hfill\it\qquad\author}
\headline={\ifnum\pageno>1{\ifodd\pageno\rightheadline\else\leftheadline\fi}
\else\fi}


\def\author{I. G. Avramidi}
\def\title{Algebraic calculation of the low-energy heat kernel}

\nopagenumbers
\null
\vskip-1.5cm
\hskip6.2cm{ \hrulefill }
\vskip-.55cm
\hskip6.2cm{ \hrulefill }
\smallskip
\hskip6.2cm{{\pl \ University of Greifswald (March, 1995)}}
\smallskip
\hskip6.2cm{ \hrulefill }
\vskip-.55cm
\hskip6.2cm{ \hrulefill }
\bigskip
\hskip6.2cm{\ hep-th/9503132}
\bigskip
\hskip6.2cm{\ submitted to:}

\hskip6.2cm{\sf \ Journal of Mathematical Physics}

\vfill

\centerline{\titlesf Covariant algebraic method for calculation of}
\medskip
\centerline{\titlesf the low-energy heat kernel}
\bigskip

\centerline{{\authorsf I. G. Avramidi}
\footnote{$^{a)}$}{ Alexander von Humboldt Fellow}
\footnote{$^{b)}$}{On leave of absence from the Research Institute for Physics,
Rostov State University, Stachki 194, Rostov-on-Don 344104, Russia}
\footnote{$^{c)}$}{E-mail: avramidi@math-inf.uni-greifswald.d400.de}}
\smallskip
\centerline{\it Department of Mathematics, University of Greifswald}
\centerline{\it Jahnstr. 15a, 17489 Greifswald, Germany}

\vfill

{\narrower
Using our recently proposed covariant algebraic approach the heat kernel for a
Laplace-like differential operator in low-energy approximation is studied.
Neglecting all the covariant derivatives of the gauge field strength
(Yang-Mills curvature) and the covariant derivatives of the potential term of
third order and higher a closed formula for the heat kernel as well as its
diagonal is obtained. Explicit formulas for the coefficients of the asymptotic
expansion of the heat kernel diagonal in terms of the Yang-Mills curvature, the
potential term and its first two covariant derivatives are obtained.
\par}
\eject


\leftline{\subchaptsf I. INTRODUCTION}
\bigskip

The heat kernel, the kernel of the one-parameter semigroup (or the heat {\it
operator}), $U(t)=\exp(-t H)$, for an elliptic differential operator $H$ acting
on a section of a vector bundle over a manifold $M$, plays a very important
role in various areas of mathematical physics, especially in quantum field
theory and quantum gravity.$^{1-28}$ It determines among others such
fundamental objects of the quantum field theory as the Green function, the
kernel of the resolvent$^2$
$$
G(\l)=(H+\l)^{-1}=\int\limits_0^\infty dt e^{-t\l} U(t),
\eqno(1.1)
$$
the zeta-function,(see Ref. 15),
$$
\zeta(p)=\Tr H^{-p}={1\over\Gamma(p)}\int\limits_0^\infty dt t^{p-1}\Tr U(t),
\eqno(1.2)
$$
where $\Tr$ is the functional trace, the functional determinant, $\Det H$,
and, hence, the one-loop effective action
$$
\Gamma_{(1)}={\frac 1 2}\log\Det H = -{\frac 1 2} \zeta'(0).
\eqno(1.3)
$$

The most important operators in quantum field theory are the second order
elliptic operators of the form
$$
H=-\sq + Q,
\eqno(1.4)
$$
where $\sq=g^{\mu\nu}\na_\mu\na_\nu$, $\na_\mu$ is a connection on the vector
bundle and $Q$ is an endomorphism of this bundle. In other words the operator
$H$ acts on a multiplet of quantum fields, $\phi(x)$, $Q(x)$ is a matrix valued
potential term, $\na_\mu=\partial_\mu+{\cal A}_\mu$, ${\cal A}_\mu$ is a gauge
(Yang-Mills) field, the gauge field strength (Yang-Mills curvature),
$\R_{\m\n}$,
being given by the commutator of covariant derivatives
$$
[\na_\mu,\na_\nu]\phi={\cal R}_{\mu\nu}\phi.
\eqno(1.5)
$$
$M$ is taken to be a $d$-dimensional Riemannian manifold $M$ with a metric,
$g_{\mu\nu}$, of Euclidean (positive) signature.

Obviously, the heat kernel is calculable {\it exactly} only in exceptional
cases of background fields configurations, (see, for example Ref. 14). That is
why for studying the general case there is a need to develop {\it approximate}
methods of calculations. Two such approximation schemes are available:$^3$ {\it
i)} the high-energy one, which corresponds to weak rapidly varying background
fields (short waves), and {\it ii)} the low-energy approximation corresponding
to the strong slowly varying background fields (long waves). The high-energy
approximation was studied in Refs. 16-19 where the heat kernel and the
effective action in second$^{16,17}$ and third$^{18,19}$ order in background
fields (curvatures) were calculated. The low-energy approximation in various
settings was studied in Refs. 20-22. The authors of these papers summed up some
particular terms in the heat kernel asymptotic expansion, such as the scalar
curvature terms$^{20,21}$ or terms without derivatives of the potential
term$^{22}$ etc.

In our recent papers$^{23-28}$ the status of the low-energy approximation in
quantum gravity and gauge theories was analyzed. We developed a new purely
algebraic {\it covariant} approach for calculating the heat kernel near
diagonal. The point is that in low-energy approximation the covariant
derivatives of the curvatures and the potential term (but not the curvature and
the potential term themselves!) are small. Therefore, one can treat them
perturbatively, the zeroth order
of this perturbation theory corresponding to the covariantly constant
background fields.

To gain greater insight into how the low-energy heat kernel looks like, one can
take into account a {\it finite} number of low-order covariant derivatives of
the background fields, and neglect all the covariant derivatives of higher
orders. Then there exist a set of differential operators, actually approximate
Killing vectors, that together with the background fields and their low-order
derivatives generate a {\it finite} dimensional Lie algebra. If one does not
neglect the higher derivatives of the background fields then this algebra is
{\it infinite}-dimensional.
This procedure is very similar to the polynomial approximation of functions of
real variables. The difference is that we are dealing, in general, with the
{\it covariant} derivatives and the curvatures.

In the previous papers we considered the case of covariantly constant gauge
field strength and the potential term in flat space, $\na_\m\R_{\a\b}=\na_\m
Q=R_{\a\b\g\d}=0$,$^{23-25}$ and the case of covariantly constant Riemann
curvature and the potential term without the Yang-Mills curvature, $\na_\m
R_{\a\b\g\d}=\na_\m Q=\R_{\m\n}=0$,$^{26,27}$. In the Ref. 28 this method was
applied for the investigation of the effective potential of a non-Abelian gauge
theory.

In this paper we are going to take into account the next terms in the
low-energy approximation in the flat space, i.e. we consider the case $\na_\m
\R_{\a\b}=\na_\m\na_\n\na_\l Q=0$, $R_{\a\b\g\d}=0$, and take into account the
first and the second derivatives of the potential term in the presence of the
covariantly constant Yang-Mills curvature.


\bigskip
\bigskip
\leftline{\subchaptsf II. COVARIANT ALGEBRAIC APPROACH}
\bigskip
\leftline{\subchaptsf A. Low-energy Lie algebra}
\bigskip

As we will study only {\it local} effects in the low-energy approximation, we
will not take care about the topology of the manifold $M$. To be precise one
can take, for example, $\RR^d$. We assume only that in {\it some region} of the
manifold the background fields satisfy approximately the following low-energy
local conditions
$$
R_{\mu\nu\a\b}=0, \qquad \na_\a {\cal R}_{\mu\nu}=0, \qquad
\na_\mu\na_\nu\na_\l Q=0.
\eqno(2.1)
$$
Commuting the covariant derivatives here leads to the following restrictions
$$
[{\cal R}_{\a\b}, {\cal R}_{\mu\nu}]=0, \qquad
[{\cal R}_{\a\b}, Q_{;\nu}]=0, \qquad
[{\cal R}_{\a\b}, Q_{;\mu\nu}]=0, \qquad
[{\cal R}_{\a\b}, [{\cal R}_{\mu\nu}, Q]]=0,
\eqno(2.2)
$$
where $Q_{;\mu}\equiv\na_\mu Q$, $Q_{;\nu\mu}\equiv\na_\mu\na_\nu Q$.
Although the conditions (2.2) do not imply directly
$$
[{\cal R}_{\a\b}, Q]=0,
\eqno(2.3)
$$
we assume (2.3) to be true since this follows from (2.2) if we suppose that
(2.2) should be valid for arbitrary gauge group. Therefore, we assume,
actually, the background to be approximately {\it Abelian}, i.e. all the
nonvanishing background quantities, ${\cal R}_{\a\b}$, $Q$, $Q_{;\mu}$,
$Q_{;\nu\mu}$, are assumed to commute with each other, in other words, to lie
in the Cartan subalgebra of the algebra of the gauge group, so that in addition
to (2.2) and (2.3) it holds
$$
[Q, Q_{;\mu}]=0, \qquad [Q, Q_{;\mu\nu}]=0, \qquad [Q_{;\mu}, Q_{;\a\b}]=0.
\eqno(2.4)
$$
For our purposes, it is helpful to introduce the following parametrization of
the potential term
$$
Q=M-\b^{ik}L_iL_k,
\eqno(2.5)
$$
where $(i=1,\dots,p)$, $p\le d$, $\b^{ik}$ is some constant symmetric
nondegenerate $p\times p$ matrix, $M$ is a covariantly constant matrix and
$L_i$ are some matrices with vanishing {\it second} covariant derivative
$$
\na_\m M=0, \qquad \na_\m\na_\n L_i=0.
\eqno(2.6)
$$
The derivatives of the potential term take then the form
$$
Q_{;\m}=-2E_{\m i}\b^{ik}L_k, \qquad
Q_{;\n\m}=2P_{\n\m}
\eqno(2.7)
$$
where
$$
E_{\mu i}=\na_\m L_i, \qquad P_{\m\n}=-E_{\m k}\b^{ki}E_{\n i}.
\eqno(2.8)
$$

Thus we have a nilpotent Lie algebra, $\{\na_\mu,$ ${\cal R}_{\a\b},$ $M,$
$L_i,$ $E_{\m i}\}$, with following nontrivial commutators
$$
[\na_\mu, \na_\nu]={\cal R}_{\mu\nu}, \qquad
[\na_\mu, L_i] = E_{\mu i},
\eqno(2.9)
$$
and the center $\{{\cal R}_{\a\b},$ $M,$ $L_i,$ $E_{\mu i}\}$.
Introducing the generators $X_A=(\na_\m, L_i)$, $(A=1,\dots, D)$, $D=d+p$, one
can rewrite these commutation relations in a more compact form
$$
[X_A, X_B]=\F_{AB}, \qquad
\eqno(2.10)
$$
$$
[X_A, \F_{CD}]=[\F_{AB}, \F_{CD}]=0,
$$
where $\F_{AB}$ is a matrix
$$
(\F_{AB})=\left(\matrix{
\R_{\m\n} & E_{\m i}\cr
-E_{\n k} & 0       \cr}\right),
\eqno(2.11)
$$
that we call the {\it generalized} curvature.
The operator $H$ (1.4) can now be written in the form
$$
H=-\g^{AB}X_A X_B+M,
\eqno(2.12)
$$
where
$$
(\g^{AB})=\left(\matrix{g^{\m\n} & 0 \cr
			0 & \b^{ik} \cr}\right).
\eqno(2.13)
$$
The matrices $\b^{ik}$ and $\g^{AB}$ play the role of metrics and can be used
to raise and to lower the small and the capital Latin indices respectively.


\bigskip
\bigskip
\leftline{\subchaptsf B. Analytic functions of the generalized curvature}
\bigskip

Let us now introduce the following matrix notation for the generalized
curvature
$$
\F=(\F^A_{\ B})=\left(\matrix{\R & E\cr
-\bar E &0\cr
}\right),
\eqno(2.14)
$$
$$
\R=(\R^\m_{\ \n}), \qquad E=(E^\m_{\ i}), \qquad \bar E=(\bar E^k_{\
\n})=(E_\n^{\ k}),
$$
and consider analytic functions of this matrix.

The powers of the matrix $\F$ (2.14) have the following general structure
$$
\F^0=1,\qquad
\F^k=\left(\matrix{J_k & J_{k-1}E \cr
-\bar EJ_{k-1} & -\bar EJ_{k-2}E\cr}\right), \qquad k\ge 1,
\eqno(2.15)
$$
where the $d\times d$ matrices $J_k=(J_{k\ \ \n}^{\ \m})$ are defined by the
recursion relations
$$
J_{k+1}=\R J_k+P J_{k-1},
\eqno(2.16)
$$
with the following initial conditions
$$
J_{-1}=0, \qquad J_0=1.
\eqno(2.17)
$$
The low-order matrices are
$$
\eqalignno{
J_0&=1, &\cr
J_1&=\R, &\cr
J_2&=\R^2+P, &(2.18)\cr
J_3&=\R^3+\R P+P\R, &\cr
J_4&=\R^4+\R^2P+\R P\R +P\R^2+P^2, &\cr}
$$
etc. It is clear that $J_k$ is a polynomial of two noncommuting `variables'
$\R$ and $P$ with all possible terms of the dimension $\R^k$ with coefficients
+1,
the total number, $N_k$, of terms being
$$
N_k=\sum_{0\le n \le [k/2]}{k-n\choose n}.
\eqno(2.19)
$$
The polynomials $J_k$ can be calculated with the help of the generating
function
$$
F(z)=(1-z\R-z^2P)^{-1}=\sum_{k\ge 0}z^k J_k,
$$
in the following way
$$
J_k={1\over k!}{\partial^k\over \partial z^k}F(z)\Big\vert_{z=0}
=\oint\limits_C {d z\over 2\pi i} z^{-k-1}F(z),
\eqno(2.21)
$$
where the integral is taken along a sufficiently small closed contour $C$ that
encircles the origin counter-clockwise, so that $F(z)$ is analytic inside this
contour.

Using the symmetry properties of the matrices $\R$ and $P$
$$
\R^T=-g\R g^{-1}, \qquad P^T=gPg^{-1},
\eqno(2.22)
$$
where $g=(g_{\m\n})$, we find
$$
J_{2k}^T=gJ_{2k}g^{-1}, \qquad J_{2k+1}^T=- gJ_{2k+1}g^{-1},
\eqno(2.23)
$$
and
$$
(\F^{2k})^T=\g\F^{2k}\g^{-1}, \qquad (\F^{2k+1})^T=-\g\F^{2k+1}\g^{-1},
\eqno(2.24)
$$
where $\g=(\g_{AB})$.
Therefore, the traces of odd order polynomials as well as the traces of odd
powers of $\F$ vanish
$$
\tr J_{2k+1}=0, \qquad
\tr\F^{2k+1}=0.
\eqno(2.25)
$$
Hence using (2.15) we obtain the traces of even powers
$$
\tr\F^{2k}=\tr(J_{2k}+PJ_{2k-2}).
\eqno(2.26)
$$

Thus any analytic function of the matrix $\F$ has the form
$$
\eqalignno{
f(\F)=&\sum\limits_{n\ge 0}{1\over n!}f^{(n)}\F^n
=\left(\matrix{U & VE \cr
-\bar EV & f(0)-\bar EWE\cr}\right)&\cr
=&\oint\limits_C{d z \over 2\pi i}{1\over z}
f\left(z^{-1}\right)(1-z\F)^{-1},&(2.27)\cr}
$$
where $U=(U^\m_{\ \n})$ and $W=(W^\m_{\ \n})$ are symmetric and $V=(V^\m_{\
\n})$ antisymetric $d\times d$ matrices
$$
U^T=gUg^{-1}, \qquad W^T=gWg^{-1}, \qquad V^T=-gVg^{-1},
\eqno(2.28)
$$
defined by
$$
\eqalignno{
U&=\sum_{n\ge 0}{1\over n!}f^{(n)}J_n
=\oint\limits_C{d z \over 2\pi i}{1\over z}f({z^{-1}})F(z),&(2.29)\cr
V&=\sum_{n\ge 1}{1\over n!}f^{(n)}J_{n-1}
=\oint\limits_C{d z \over 2\pi i}f(z^{-1})F(z),&(2.30)\cr
W&=\sum_{n\ge 2}{1\over n!}f^{(n)}J_{n-2}
=\oint\limits_C{d z \over 2\pi i}zf(z^{-1})F(z).&(2.31)\cr}
$$
Herefrom we find the trace of an analytic function, $f(\F)$ (2.27),
$$
\eqalignno{
\tr f(\F)=&\tr(U+PW)
=\tr f(0)+\sum_{n\ge 2}{1\over n!}f^{(n)}\tr(J_n+PJ_{n-2})&\cr
=&\oint\limits_C{d z \over 2\pi i}f(z^{-1})\tr
\left\{(z^{-1}+zP)F(z)\right\},&(2.32)\cr}
$$
and the determinant
$$
\det\left({f(\F)\over f(0)}\right)
=\exp\left\{\oint\limits_C{d z \over 2\pi i}
\log\left({f(z^{-1})\over f(0)}\right)\tr
\left\{(z^{-1}+zP)F(z)\right\}\right\}.
\eqno(2.33)
$$

These formulas take especially simple form in a particular case when the
matrices $\R$ and $P$ commute,
$$
[\R, P]=0.
\eqno(2.34)
$$
In this case the generating function $F(z)$ (2.20) can be evaluated exactly
$$
F(z)={1\over \Delta}\left({R_+\over 1-zR_+}-{R_-\over 1-zR_-}\right),
\eqno(2.35)
$$
where
$$
R_\pm=\frac 1 2(\R\pm\Delta), \qquad
\Delta=\sqrt{\R^2+4P}.
\eqno(2.36)
$$
Thereby we find
$$
J_k={1\over\Delta}(R_+^{k+1}-R^{k+1}_-).
\eqno(2.37)
$$
Using these polynoms we obtain from (2.29)-(2.31)
$$
\eqalignno{
U&=\Delta^{-1}(R_+f(R_+)-R_-f(R_-))&\cr
&=\frac 1 2(f(R_+)+f(R_-))+\frac 1 2\R V,&(2.38)\cr
V&=\Delta^{-1}(f(R_+)-f(R_-)), &(2.39)\cr
W&=\Delta^{-1}(R_+^{-1}f(R_+)-R_-^{-1}f(R_-)),&(2.40)\cr}
$$
and, consequently, from (2.32) and (2.33)
$$
\tr f(\F)=\tr(f(R_+)+f(R_-)).
\eqno(2.41)
$$
$$
\det\left({f(\F)\over f(0)}\right)=\det\left({f(R_+)\over f(0)}{f(R_-)\over
f(0)}\right).
\eqno(2.42)
$$

\bigskip
\leftline{\subchaptsf C. The heat operator}
\bigskip

Now using the algebra (2.10) we are able to calculate the heat kernel operator,
$\exp(-tH)$. For an algebra of this kind in Refs. 23,24 it was proven a theorem
that gives the heat operator in terms of an integral over the corresponding Lie
group. Namely,
$$
\eqalignno{
\exp(-t H)=&(4\pi t)^{-D/2}
\det\left({\sinh(t\F)\over t\F}\right)^{-1/2}\exp(-t M)
&\cr&\times\int\limits_{\RR^D} d k \g^{1/2}
\exp\left\{-{1\over 4t}k_A(t\F \coth(t\F))^A_{\ B}k^B +
k^A X_A\right\},&(2.43)\cr}
$$
where $\g=\det\g_{AB}$, $(\g_{AB})=(\g^{AB})^{-1}$.

Thus we have expressed the heat kernel operator in terms of the operator
$\exp(k^A X_A)$.
Splitting the integration variables $(k^A)=(q^\m, \om^i)$ in (2.43) and using
the Campbell-Hausdorf formula we single out the noncommutative part
$$
\eqalignno{
\exp(k^AX_A)=\exp(q^\m\na_\m+\om^iL_i)
&=\exp(\om^iL_i)\exp(\frac 1 2 [q^\m\na_\m, \om^iL_i])
\exp(q^\m\na_\m)&\cr
&=\exp\{\om^i(L_i+\frac 1 2 q^\m E_{\m i})\}
\exp(q^\m\na_\m).&(2.44)\cr}
$$
Further, we employ the parametrization
$$
t\F \coth(t\F)=\left(\matrix{
B(t) & tA(t)E\cr
-t\bar EA(t) & 1-t^2\bar E C(t) E \cr
}\right),
\eqno(2.45)
$$
and integrate over $\om$ in (2.43) to yield
$$
\eqalignno{
\exp(-t H)=&(4\pi t)^{-d/2}\det\left({\sinh(t\F)\over t\F}\right)^{-1/2}
\det Z(t)^{1/2}
\exp\left\{-t M+tL_iZ^i_{\ k}(t)L^k\right\} &\cr
&\times\int\limits_{\RR^d} d q\, g^{1/2}\exp\left\{
-{1\over 4t}q^\m D_{\m\n}(t)q^\n
+tL_iY^i_{\ \m}(t)q^\m \right\}
\exp(q^\m\na_\m),&(2.46)\cr}
$$
where
$$
Z(t)=(Z^i_{\ k}(t))=(1-t^2\bar EC(t)E)^{-1},
\eqno(2.47)
$$
$$
Y(t)=(Y^i_{\ \m}(t))=Z(t)\bar E(1+A(t)),
\eqno(2.48)
$$
$$
D(t)=(D^\m_{\ \n}(t))=B(t)-t^2(1-A(t))EZ(t)\bar E(1+A(t)).
\eqno(2.49)
$$
Further, using (2.5) and (2.7) it is not difficult to prove
$$
L_iY^i_{\ \n}(t)=
-\frac 1 2 Q_{;\m}(1+t^2CP)^{-1 \m}_{\ \ \ \n}
\eqno(2.50)
$$
$$
L_iZ^{i}_{\ k}(t)L^k=M-Q+ \frac 1 4 t^2 Q_{;\m}
[(1+t^2CP)^{-1}C]^{\m}_{\ \n}Q^{;\n},
\eqno(2.51)
$$
$$
\det Z(t)=\det(1+t^2CP)^{-1},
\eqno(2.52)
$$
where $P=(P^\m_{\ \n})=-E\bar E$ is given by (2.8).
Substituting these expressions in (2.46) we arrive finally to the heat operator
in the form
$$
\eqalignno{
\exp(-&tH)=(4\pi t)^{-d/2}\exp\left\{-t Q + \Phi(t)
+ \frac 1 4 t^3 Q_{;\m}\Psi^{\m\n}(t)Q_{;\n}\right\} &\cr
&\times\int\limits_{\RR^d} d q\, g^{1/2}
\exp\left\{-{1\over 4t}q^\m D_{\m\n}(t)q^\n
- {t\over 2}Q_{;\m}(\d^\m_\n+A^\m_{\ \n}(t))q^\n\right\}
\exp(q^\m\na_\m), &(2.53)\cr}
$$
where
$$
D(t)=(D^\m_{\ \n}(t))=B(t)+t^2(1-A(t))P(1+t^2C(t)P)^{-1}(1+A(t)),
\eqno(2.54)
$$
$$
\Phi(t)=-\frac 1 2 \log\det\left({\sinh(t\F)\over t\F}\right)
-\frac 1 2 \log\det(1+t^2C(t)P),
\eqno(2.55)
$$
$$
\Psi(t)=(\Psi^\m_{\ \n}(t))=(1+t^2C(t)P)^{-1}C(t),
\eqno(2.56)
$$
the matrices $A(t)$, $B(t)$ and $C(t)$ being defined from (2.29)-(2.31) and
(2.45) by
$$
\eqalignno{
B(t)=&\oint\limits_C{dz\over 2\pi i}{t\over z^2}\coth({tz^{-1}})F(z),
&(2.57)\cr
A(t)=&\oint\limits_C{dz\over 2\pi i}{t\over z}\coth({tz^{-1}})F(z), &(2.58)\cr
C(t)=&\oint\limits_C{dz\over 2\pi i}t \coth({tz^{-1}})F(z), &(2.59)\cr}
$$
with $F(z)$ given by (2.20).
It is worth noting that in the parametrization
$$
{\sinh(t\F)\over t\F}
=\left(\matrix{
K(t) & tS(t)E\cr
-t\bar ES(t) & 1-t^2\bar E N(t) E \cr
}\right),
\eqno(2.60)
$$
where
$$
\eqalignno{
K(t)=&\oint\limits_C{dz\over 2\pi i}{t\over z^{2}}\sinh(tz^{-1})F(z),
&(2.61)\cr
S(t)=&\oint\limits_C{dz\over 2\pi i}{t\over z}\sinh(tz^{-1})F(z), &(2.62)\cr
N(t)=&\oint\limits_C{dz\over 2\pi i}t\,\sinh(tz^{-1})F(z), &(2.63)\cr}
$$
one can calculate the determinant of this $D\times D$ matrix as follows
$$
\eqalignno{
\det \left({\sinh(t\F)\over t\F}\right)
&=\det K\det\left[1+t^2(N-SK^{-1}S)P\right]&\cr
&=\det(1+t^2 NP)\det\left[K-t^2SP(1+t^2NP)^{-1}S\right].&(2.64)\cr}
$$

\bigskip
\bigskip
\leftline{\subchaptsf III. THE HEAT KERNEL}
\bigskip
\leftline{\subchaptsf A. Closed formula for the heat kernel diagonal}
\bigskip

To obtain the heat kernel in coordinate representation we have just to act with
the heat operator, $\exp(-t H)$, on the coordinate $\d$-function
$$
U(t|x,x')=\exp(-t H)\P(x,x')g^{-1/2}\d(x-x'),
\eqno(3.1)
$$
where $\P(x,x')$ is the parallel displacement operator (matrix) of the field
$\phi$ from the point $x'$ to the point $x$ along the geodesic.
It is not difficult to show$^{23,24}$ that
$$
\exp(q^\m\na_\m^x)\P(x,x')g^{-1/2}\d(x-x')=\P(x,x')g^{-1/2}\d(x-x'+q).
\eqno(3.2)
$$
Hence the integration over $q$ in (2.53) becomes trivial and we obtain the heat
kernel
$$
\eqalignno{
U(t|&x,x')=(4\pi t)^{-d/2}\P(x,x')\exp\left\{-t Q(x) + \Phi(t)
+ \frac 1 4 t^3 Q_{;\m}(x)\Psi^{\m\n}(t)Q_{;\n}(x)\right\} &\cr
&\times\exp\left\{-{1\over 4t}(x-x')^\m D_{\m\n}(t)(x-x')^\n
+{t\over 2}Q_{;\m}(x)(\d^\m_\n+A^\m_{\ \n}(t))(x-x')^\n\right\}.&(3.3)\cr}
$$
Expanding this expression in a power series in $(x-x')$ one can easily get {\it
all} the coincidence limits of covariant derivatives of the heat kernel. In
particular, the heat kernel diagonal has a very simple form
$$
[U(t)]=U(t|x,x)=(4\pi t)^{-d/2}\exp\left\{-t Q + \Phi(t)
+ \frac 1 4 t^3 Q_{;\m}\Psi^{\m\n}(t)Q_{;\n}\right\}.
\eqno(3.4)
$$
This is the main result of this paper. The formula (3.4) exhibits the general
structure of the heat kernel diagonal. Namely, one sees immediately how the
potential term and its first derivatives enter the result. The complete
nontrivial information is contained only in a scalar, $\Phi(t)$, and a tensor,
$\Psi_{\m\n}(t)$, functions which are constructed purely from the Yang-Mills
curvature $\R_{\m\n}$ and the {\it second} derivatives of the potential term,
$\na_\m\na_\n Q$. So we conclude that the coefficients of the heat kernel
asymptotic expansion are constructed from three different types of scalar
(connected) blocks, $Q$, $\Phi_{(n)}(\R, \na\na Q)$ and $\na_\m
Q\Psi^{\m\n}_{(n)}(\R, \na\na Q)\na_\n Q$. We will calculate the coefficients
of the heat kernel asymptotic expansion explicitly in the Subsects. III.B and
III.C.

Before we do this let us consider the particular case (2.34) when the matrices
$\R$ and $P$ commute, i.e.
$$
\R^\m_{\ \n}P^\n_{\ \a}=P^\m_{\ \n}\R^\n_{\ \a}.
\eqno(3.5)
$$
Making use of the formulas (2.45) and (2.36)-(2.42) we get
$$
\det\left({\sinh(t\F)\over t\F}\right)=\det\left({\sinh(tR_+)\over
tR_+}{\sinh(tR_-)\over tR_-}\right),
\eqno(3.6)
$$
$$
C(t)=-{1\over t^2 P}-{\sinh(t\Delta)\over t\Delta}{1\over
\sinh(tR_+)\sinh(tR_-)}.
\eqno(3.7)
$$
Now from (2.55) and (2.56) it follows
$$
\Phi(t)=-\frac 1 2 \log\det\left({\sinh(t\Delta)\over t\Delta}\right),
\eqno(3.8)
$$
$$
\Psi(t)={1\over t^2 P}
\left[{\Delta\over 2tP}{\cosh(t\R)-\cosh(t\Delta)\over
\sinh(t\Delta)}+1\right].
\eqno(3.9)
$$
Thus in this special case the heat kernel diagonal reads
$$
\eqalignno{
[U(t)]&=(4\pi t)^{-d/2}\det\left(\sinh(t\Delta)\over t\Delta\right)^{-1/2}&\cr
&\times\exp\left\{-t Q + \frac 1 4 t Q_{;\m}
\left[{1\over P}
\left({\Delta\over 2tP}{\cosh(t\R)-\cosh(t\Delta)\over \sinh(t\Delta)}
+1\right)\right]^\m_{\ \n}
Q^{;\n}\right\}. &(3.10)\cr}
$$

If the second derivatives of the potential vanish, $P_{\m\n}=\frac 1 2
\na_\m\na_\n Q=0$, then we have from (3.8)-(3.10)
$$
\Phi(t)=-\frac 1 2 \log\det\left({\sinh(t\R)\over t\R}\right),
\eqno(3.11)
$$
$$
\Psi(t)={1\over t^2\R^2}(t\R\coth(t\R)-1).
\eqno(3.12)
$$
The heat kernel diagonal is now
$$
\eqalignno{
[U(t)]=&(4\pi t)^{-d/2}\det\left(\sinh(t\R)\over t\R\right)^{-1/2}&\cr
&\times
\exp\left\{-t Q + \frac 1 4 t Q_{;\m}
\left({{1\over \R^2}(t\R\coth(t\R)-1)}\right)^\m_{\ \n}
Q^{;\n}\right\}. &(3.13)\cr}
$$
This is the first order correction to the case of covariantly constant
potential$^{23}$ when additionally the {\it first} derivatives of the potential
are taken into account.

In the case of vanishing Yang-Mills curvature, $\R=0$, we have similarly
$$
\Phi(t)=-\frac 1 2 \log\det\left({\sinh(2t\sqrt P)\over 2t\sqrt P}\right),
\eqno(3.14)
$$
$$
\Psi(t)=-{1\over (t\sqrt P)^3}(\tanh(t\sqrt P)-t\sqrt P),
\eqno(3.15)
$$
and the heat kernel diagonal has the form
$$
\eqalignno{
[U(t)]=&(4\pi t)^{-d/2}\det\left(\sinh(2t\sqrt P)\over 2t\sqrt P\right)^{-1/2}
&\cr&
\times\exp\left\{-t Q
- \frac 1 4 Q_{;\m}
\left({\tanh(t\sqrt P)-t\sqrt P\over P^{3/2}}\right)^\m_{\ \n}
Q^{;\n}\right\}. &(3.16)\cr}
$$
This determines the low-energy approximation without the gauge fields.
The formulas (3.10), (3.13) and (3.16) can be used, in particular, to check the
results for the coefficients of the heat kernel asymptotic expansion obtained
in general case.


\bigskip
\bigskip
\leftline{\subchaptsf B. Asymptotic expansion of the heat kernel diagonal}
\vglue0pt
\bigskip
\vglue0pt
Let us now calculate the Taylor expansions of the functions $\Phi(t)$ (2.55)
and $\Psi(t)$ (2.56).
Using the well known series$^{29}$
$$
\log{\sinh x\over x}=\sum_{n\ge 1}{2^{2n-1}B_{2n}\over n (2n)!}x^{2n},
\eqno(3.17)
$$
where $B_n$ are the Bernulli numbers$^{29}$, we obtain from (2.32)
$$
\log\det\left({\sinh(t\F)\over t\F}\right)
=\sum_{n\ge 1}{2^{2n-1}B_{2n}\over n (2n)!}
t^{2n}\tr(J_{2n}+PJ_{2n-2}).
\eqno(3.18)
$$
Further, using another series$^{29}$
$$
x\coth x=\sum_{n\ge 0}{2^{2n}B_{2n}\over (2n)!}x^{2n},
$$
we find from (2.57)-(2.59) the functions $B(t)$, $A(t)$ and $C(t)$
$$
B(t)=\sum_{n\ge 0}{2^{2n}B_{2n}\over (2n)!}t^{2n}J_{2n},
\eqno(3.20)
$$
$$
A(t)=\sum_{n\ge 1}{2^{2n}B_{2n}\over (2n)!}t^{2n-1}J_{2n-1},
\eqno(3.21)
$$
$$
C(t)=\sum_{n\ge 1}{2^{2n}B_{2n}\over (2n)!}t^{2n-2}J_{2n-2}.
\eqno(3.22)
$$
Now it is not difficult to calculate
$$
\eqalignno{
\log\det(1+t^2C(t)P)&=\sum\limits_{k\ge 1}t^{2k}2^{2k-1}
\sum\limits_{1\le N\le k}{(-1)^N\over N}
\sum\limits_{{\scriptstyle 1\le n_1, \dots, n_N\le k\atop \scriptstyle
n_1+\cdots+n_N=k}} {B_{2n_1}\over (2n_1)!}\cdots {B_{2n_N}\over (2n_N)!}
&\cr &\times\tr(PJ_{2n_1-2}\cdots PJ_{2n_N-2}).&(3.23)\cr}
$$
Thereby combining (3.18) and (3.23) we find the function $\Phi(t)$
$$
\Phi(t)=\sum_{k\ge 1}{(-1)^k\over (2k)!}t^{2k}\Phi_{2k},
\eqno(3.24)
$$
where
$$
\eqalignno{
&\Phi_{2k}=\tr \Biggl\{{2^{2k-2}\over k}|B_{2k}|
\left(\R J_{2k-1}+2(k+1)PJ_{2k-2}\right)& (3.25)\cr
&+2^{2k-1}
\sum\limits_{2\le N\le k}{1\over N}\sum\limits_{{\scriptstyle 1\le n_1, \dots,
n_N\le k-1\atop \scriptstyle n_1+\cdots+n_N=k}}{(2k)!\over
(2n_1)!\cdots(2n_N)!}{|B_{2n_1}|\cdots |B_{2n_N}|}
PJ_{2n_1-2}\cdots PJ_{2n_N-2}\Biggr\}.&\cr}
$$
Here we used the recursion relations (2.16) and the property of the Bernulli
numbers $B_{2k}=(-1)^{k+1}|B_{2k}|$, $(k\ge 1)$.
Since all the polynoms $J_k$ have only positive coefficients we see that
$\Phi_{2k}$ are also polynoms of the matrix $\R$ and $P$ with positive
coefficients.

Similarly, from (3.22) and (2.56) we obtain the expansion of the matrix $\Psi$
$$
\Psi(t)=\sum_{k\ge 0}{(-1)^k\over (2k+3)!}t^{2k}\Psi_{2k},
\eqno(3.26)
$$
where
$$
\eqalignno{
\Psi_{2k}=&{(2k+3)2^{2k+2}|B_{2k+2}|}J_{2k}
&\cr&
+2^{2k+2}\sum\limits_{2\le N\le k+1;}\sum\limits_{{\scriptstyle 1\le n_1,
\dots, n_N\le k\atop \scriptstyle n_1+\cdots+n_N=k+1}}{(2k+3)!\over
(2n_1)!\cdots(2n_N)!}{|B_{2n_1}|\cdots |B_{2n_N}|}&\cr
&\times
J_{2n_1-2}PJ_{2n_2-2}P\cdots PJ_{2n_N-2}.&(3.27)\cr}
$$
Evidently, $\Psi_{2k}$ also is a polynom of $\R$ and $P$ having only positive
coefficients.
Analogously, using the explicit formulas (3.20)-(3.22) one can calculate the
Taylor series of the matrix $D$ (2.54) too. We will not list here the result
because it is not needed for the calculation of the heat kernel diagonal.

Finally, substituting the series (3.24) and (3.26) in (3.4) we obtain the
expansion of the heat kernel diagonal
$$
[U(t)]=(4\pi t)^{-d/2}\exp\left(\sum_{k\ge 1}(-1)^{[(k+1)/2]}{t^k\over k!}
b_k\right),
\eqno(3.28)
$$
where the coefficients $b_k$ are defined by
$$
b_1=Q,\qquad
b_{2k}=\Phi_{2k}, \qquad
b_{2k+1}=\frac 1 4 Q_{;\m}\Psi_{2k-2}^{\m\n}Q_{;\n},
\eqno(3.29)
$$
and all have only positive coefficients.
The coefficients of the usual asymptotic expansion of the heat kernel
diagonal,$^6$
$$
[U(t)]=(4\pi t)^{-d/2}\sum\limits_{k\ge 0}{(-t)^k\over k!}a_k,
\eqno(3.30)
$$
are given then by
$$
a_k=\sum\limits_{1\le N\le k;}\sum\limits_{\scriptstyle 1\le k_1\le\dots\le k_N
\le  k \atop \scriptstyle
k_1+\cdots+k_N=k}(-1)^{[(3k_1+1)/2]+\cdots+[(3k_N+1)/2]}{k!\over
(k_1!)^2\cdots(k_N!)^2}b_{k_1}\cdots b_{k_N}.
\eqno(3.31)
$$


\bigskip
\bigskip
\leftline{\subchaptsf C. Explicit results for low-order coefficients}
\bigskip

Using the formulas of the previous subsection it is not difficult  to calculate
the coefficients of the asymptotic expansion of the heat kernel diagonal
explicitly. We list below some low-order terms
$$
\eqalignno{
a_1=&b_1,&\cr
a_2=&b_1^2-b_2,&\cr
a_3=&b_1^3-3b_1b_2-b_3,&\cr
a_4=&b_1^4-6b_1^2b_2-4b_1b_3+3b_2^2+b_4,&\cr
a_5=&b_1^5-10b_1^3b_2-10b_1^2b_3+15b_1b_2^2+5b_1b_4+10b_2b_3+b_5,&(3.32)\cr
a_6=&b_1^6-15b_1^4b_2-20b_1^3b_3+15b_1^2b_4+45b_1^2b_2^2
+60b_1b_2b_3+6b_1b_5
 -15b_2b_4+10b_3^2-b_6,&\cr
a_7=&b_1^7-21b_1^5b_2-35b_1^4b_3+35b_1^3b_4+190b_1^3b_2^2+21b_1^2b_5
+210b_1^2b_2b_3-7b_1b_6
&\cr&-105b_1b_2b_4+70b_1b_3^2-21b_2b_5-35b_3b_4-b_7,&\cr
a_8=&b_1^8-28b_1^6b_2-56b_1^5b_3+70b_1^4b_4+210b_1^4b_2^2+56b_1^3b_5
+560b_1^3b_2b_3-28b_1^2b_6-420b_1^2b_2b_4
&\cr&+280b_1^2b_3^2-8b_1b_7
-168b_1b_2b_5-280b_1b_3b_4+28b_2b_6-56b_3b_5+35b_4^2+b_8,&\cr}
$$
where
$$
\eqalignno{
b_1=&Q, &\cr
b_2=&\frac 1 6 \tr(\R^2+4P),&\cr
b_4=&\frac 1 {15}\tr(\R^2+4P)^2,&\cr
b_6=&\frac 8 {63} \tr(\R^6+12\R^4P+39\R^2P^2+9\R P\R P+64P^3),&(3.33)\cr
b_8=&\frac 8 {15} \tr (\R^8+16\R^6 P+\frac {153} 3 \R^4 P^2+\frac {592} 3 \R^2
P^3
+24\R^3 P\R P
&\cr&+\frac {64} 3 \R^2 P\R^2 P+\frac {176} 3 \R P\R P^2+256 P^4),&\cr}
$$
and
$$
\eqalignno{
b_3=&\frac 1 2 Q_{;\m}Q^{;\m}, &\cr
b_5=&\frac 2 3 Q_{;\m}(\R^2+6P)^\m_{\ \n}Q^{;\n},&(3.34)\cr
b_7=&\frac 8 3 Q_{;\m}(\R^4+9\R^2P+\R P\R+\frac {51} 2 P^2)^\m_{\
\n}Q^{;\n}.&\cr}
$$
It should be reminded here that the matrix $P$ is composed of the second
derivatives of the potential term, $P=(P^\m_{\ \n})$, $P_{\m\n}=\frac 1 2
\na_\m\na_\n Q$, $\R=(\R^\m_{\ \n})$ is the Yang-Mills curvature and the trace
`tr' is taken {\it only} with respect to {\it vector} (Greek) indices, the
gauge ones being intact. For example,
$$
\tr \R^2=\R^\m_{\ \n}\R^\n_{\ \m}, \qquad
\tr P=P^\m_{\ \m}=\frac 1 2 \sq Q,
\eqno(3.35)
$$
etc. Notice that the term $\sq Q$ appears only in combination $b_2=\frac 1 6
(\R_{\m\n}\R^{\n\m}+2\sq Q)$.


\bigskip
\bigskip
\leftline{\subchaptsf IV. TRACE OF THE HEAT KERNEL}
\bigskip

Let us discuss now the functional trace of the heat kernel
$$
\Tr U(t)=\int\limits_M dx g^{1/2} \tr [U(t)].
\eqno(4.1)
$$
It is well known that if the manifold $M$ is compact then there is a {\it
classic (standard)} asymptotic expansion as $t\to 0$ $^6$
$$
\Tr U(t)\sim(4\pi t)^{-d/2}\sum\limits_{k\ge 0}t^{k/2}A_{k/2}(H,M).
\eqno(4.2)
$$
The coefficients $A_{k/2}(H,M)$ are spectral invariants of the operator $H$
which are calculable in form of integrals of local invariants of background
fields over the manifold (and the boundary). In case of manifolds without
boundary this expansion can be obtained by direct term by term integration of
the asymptotic expansion of the heat kernel diagonal (3.30) over the manifold,
i.e.
$$
A_{k}=(-1)^k k!\int\limits_M dx g^{1/2}\tr a_k, \qquad
A_{k+1/2}=0.
\eqno(4.3)
$$

For noncompact manifold the situation is far more complicated. The problem is
that the integrals like (4.3) for noncompact manifolds do not converge, in
general. This depends on the behavior of the background fields at the infinity.

Let us consider, for example, a complete noncompact manifold without boundary,
$\partial M=\emptyset$.
In the case when all background fields have only compact support or fall off
sufficiently rapidly at the infinity the invariants $A_k$ are also well defined
except
$$
A_0=\const\cdot \vol(M).
\eqno(4.4)
$$
If we introduce an auxiliary operator $H_0$ with the symbol
$H_0(x,\xi)=H_L(x_0,\xi)$, where $H_L(x_0,\xi)$ is the leading symbol of the
operator $H$ and $x_0$ is some fixed point, having trivial coefficients
$$
a_0(H_0)=1, \qquad a_k(H_0)=0, \qquad k\ge 1,
\eqno(4.5)
$$
then the trace of the difference of two heat kernels is well defined for
noncompact manifolds too,
$$
\Tr (\exp(-t H)-\exp(-tH_0))\sim (4\pi t)^{-d/2}\sum\limits_{k\ge
1}t^{k}A_{k}(H,M).
\eqno(4.6)
$$
This case is similar to the {\it standard} compact one.

However, if the background fields (curvature or the potential term) do not
decrease at the infinity then the asymptotics of the trace of the heat kernel
changes radically. In this case  {\it all} the coefficients $A_k$ (4.3) diverge
and. This means that the trace of the heat kernel has a different {\it
nonstandard} asymptotics. One can show that the main term of the asymptotics
can be {\it always} obtained by
$$
\eqalignno{
\Tr U(t)\Bigg\vert_{t\to 0}&\sim (2\pi)^{-d}\tr\int\limits_M dx
\int\limits_{\RR^d}d\xi\exp(-tH(x,\xi))\Bigg\vert_{t\to 0}&\cr
&\sim (4\pi t)^{-d/2}\tr\int\limits_M dx g^{1/2}\exp(-tQ(x))\Bigg\vert_{t\to
0},&(4.7)\cr}
$$
where $H(x,\xi)$ is the symbol of the operator $H$ (1.4).

Therefore, the asymptotics depends essentially on the behavior of the potential
term at the infinity. If the potential is positive definite and increases at
the infinity then the integral over the manifold always exist and determines
the main term of the asymptotics. For example, if the potential at the infinity
behaves like
$$
Q(x)\Big\vert_{x\to \infty}=P_{\m\n}x^\m x^\n+O(x),
\eqno(4.8)
$$
and the matrix $P=(P^\m_{\ \n})$ is nondegenerate, then we obtain from (4.7)
asymptotically
$$
\Tr U(t)\sim (2t)^{-d}\det P^{-1/2}+O(t^{-d+1}).
\eqno(4.9)
$$
This is a typical example of a {\it nonstandard} asymptotics. Instead of usual
behavior, $\sim t^{-d/2}$, we have a more singular one, $\sim t^{-d}$.
Generally, there can be far more complicated asymptotics, including the
logarithmic terms, $(\log t)^\a$, etc., depending on the the asymptotic
behavior of the potential term at the infinity.

Let us now calculate the trace of the heat kernel in our case. We assumed that
the background fields satisfy the low-energy conditions (2.1)-(2.4) in some
region of the manifold $M$. Let us suppose the manifold $M$ to be $\RR^d$ and
the conditions (2.1)-(2.4) to hold everywhere. Then the formula for the heat
kernel diagonal (3.4) is valid everywhere too. If we write it in the form
$$
\eqalignno{
[U(t)]=&(4\pi t)^{-d/2}\det\left(\sinh(t\F)\over t\F\right)^{-1/2}
\det(1+t^2C(t)P)^{-1/2}&\cr
&\times\exp\left\{-t M -t(x-x_0)^\m \Pi_{\m\n}(t)(x-x_0)^\n\right\},&(4.10)\cr}
$$
where $x_0$ is some fixed point and
$$
\Pi(t)=(\Pi^\m_{\ \n}(t))=P(1+t^2C(t)P)^{-1},
\eqno(4.11)
$$
and suppose the matrix $P$ to be nondegenerate, then one can integrate (4.10)
over $\RR^d$ to get
$$
\Tr U(t)=(2t)^{-d}\det\left(\sinh(t\F)\over t\F\right)^{-1/2}\det
P^{-1/2}\exp(-tM).
\eqno(4.12)
$$
This expression also has a {\it nonclassic} asymptotics, $\Tr
U(t)\sim\const\cdot t^{-d}$.
In particular case of commuting matrices $\R$ and $P$ the trace of the heat
kernel takes especially simple form
$$
\Tr U(t)=\det\{2(\cosh(t\Delta)-\cosh(t\R))\}^{-1/2}\exp(-tM),
\eqno(4.13)
$$
which reduces to
$$
\Tr U(t)=\det(2\sinh(t\sqrt P))^{-1}\exp(-tM),
\eqno(4.14)
$$
when $\R=0$.

It should be noted that the {\it standard} form of the asymptotics of the trace
of the heat kernel (4.2) is the basis for the regularization and
renormalization procedure in quantum field theory.$^2$ That is why the
non-standard asymptotics may cause serious technical problems in the theory of
quantum fields on noncompact manifolds with background fields that do not fall
off at infinity. For example, the analytical structure of the zeta function
(1.2) in non-standard case will be completely different. This is the
consequence of the fact that in this non-standard situation the physical
quantum states are not well defined.


\bigskip
\bigskip
\leftline{\subchaptsf V. CONCLUSION}
\bigskip

In present paper we continued the investigation of the heat kernel in
low-energy approximation initiated in our recent papers.$^{23-28}$ We developed
further a manifestly covariant algebraic approach for calculation of the heat
{\it operator}, i.e. the one-parameter semigroup, $\exp(-t H)$, proposed in
these papers and applied it to evaluate the heat {\it kernel} and, especially,
the heat kernel diagonal. We were able to take into account in the asymptotic
expansion of the heat kernel diagonal {\it all} the terms that contain only the
Yang-Mills curvature, $\R_{\m\n}$, the potential term $Q$ and its two low-order
derivatives, $\na_\m Q$ and $\na_\m\na_\n Q$.

We obtained a closed formula for the heat kernel diagonal that can be treated
as a resummation of the asymptotic expansion, those terms being summed up
first. The covariant algebraic approach employed in this paper is especially
adequate and effective to study the low-energy approximation. It seems that it
can be developed deeper and that it can be formulated a general technique for
systematic calculation of the low-energy heat kernel, a kind of {\it low-energy
covariant perturbation theory}.


\bigskip
\bigskip
\leftline{\subchaptsf ACKNOWLEDGEMENTS}
\vglue0pt
\bigskip
\vglue0pt
I would like to thank R. Schimming  for many helpful discussions and J.
Eichhorn for the hospitality at the
University of Greifswald. This work was supported, in part, by the Alexander
von Humboldt Foundation.

\bigskip
\bigskip

\item{$^1$} J. S. Schwinger, Phys. Rev. {\bf 82}, 664 (1951).
\item{$^2$} B. S. De Witt, {\it Dynamical theory of groups and fields} (Gordon
and Breach, New York, 1965);
Phys. Rep. C {\bf 19}, 296 (1975);
in {\it General Relativity}, edited by S. Hawking and W. Israel (Cambridge
University Press, Cambridge, 1979).
\item{$^3$} G. A. Vilkovisky, in {\it Quantum Gravity}, edited by S.
Christensen (Adam Hilger, Bristol, 1983), p. 169.
\item{$^4$} A. O. Barvinsky and G. A. Vilkovisky, Phys. Rep. C {\bf 119}, 1
(1985).
\item{$^5$} I. G. Avramidi, Nucl. Phys. B {\bf 355}, 712 (1991); Phys. Lett. B
{\bf 238}, 92 (1990);
Teor. Mat. Fiz. {\bf 79}, 219 (1989).
\item{$^6$} P. B. Gilkey, J. Diff. Geom. {\bf 10}, 601 (1975); {\it Invariance
theory, the heat equation and the Atiyah - Singer index theorem} (Publish or
Perish, Wilmington, 1984).
\item{$^7$} R. Schimming, Beitr. Anal. {\bf 15}, 77 (1981);
Math. Nachr. {\bf 148}, 145 (1990).
\item{$^8$} R. Schimming, in {Analysis, Geometry and Groups: A Riemann Legacy
Volume}, edited by H. M. Srivastava and Th. M. Rassias, (Hadronic Press, Palm
Harbor, 1993), part. II, p. 627.
\item{$^{9}$} S. A. Fulling, SIAM J. Math. Anal. {\bf 14}, 780 (1983).
\item{$^{10}$} T. A. Osborn and F. H. Molzahn, Phys. Rev. A {\bf 34}, 1696
(1986).
\item{$^{11}$} F. H. Molzahn, T. A. Osborn and S. A. Fulling, Ann. Phys. (N.Y.)
{\bf 204}, 64 (1990);
Ann. Phys. (N.Y.) {\bf 214}, 102 (1992).
\item{$^{12}$} F. H. Molzahn and T. A. Osborn, Ann. Phys. (N. Y.) {\bf 230},
343 (1994).
\item{$^{13}$} I. G. Avramidi and R. Schimming, {\it The heat kernel
coefficients to the matrix Schr\"odinger operator}, University of Greifswald
(1995), hep-th/9501026.

\item{$^{14}$} R.  Camporesi, Phys. Rep. {\bf 196}, 1 (1990).
\item{$^{15}$} E. Elizalde, S. D. Odintsov, A. Romeo, A. A. Bytsenko and S.
Zerbini, {\it Zeta regularization techniques with applications} (World
Scientific, Singapore, 1994).

\item{$^{16}$} I. G. Avramidi, Phys. Lett. B {\bf 236}, 443 (1990); Yad. Fiz.
49 (1989) 1185.
\item{$^{17}$} A. O. Barvinsky and G. A. Vilkovisky, Nucl. Phys. B {\bf 282},
163 (1987); Nucl. Phys. B {\bf 333}, 471 (1990).
\item{$^{18}$} G. A. Vilkovisky, in {\it Publication de l' Institut de
Recherche Math\'ematique  Avanc\'ee}, (R. C. P. {\bf 43} Strasbourg, 1992), p.
203.
\item{$^{19}$} A. O. Barvinsky, Yu. V. Gusev, G. A. Vilkovisky and V. V.
Zhytnikov, J. Math. Phys. {\bf 35}, 3525 (1994);
{\it Covariant perturbation theory (IY)}, University of Manitoba (1993).

\item{$^{20}$} L. Parker and D. J. Toms, Phys. Rev. D {\bf 31}, 953 (1985).
\item{$^{21}$} I. Jack and L. Parker, Phys. Rev. D {\bf 31}, 2439 (1985).
\item{$^{22}$} J. A. Zuk, Phys. Rev. D {\bf 34}, 1791 (1986);
Phys. Rev.  D {\bf 33}, 3645 (1986).

\item{$^{23}$} I. G. Avramidi, Phys. Lett. B {\bf 305}, 27 (1993).
\item{$^{24}$} I. G. Avramidi, {\it Covariant methods for calculating the
low-energy effective action in quantum field theory and quantum gravity},
University of Greifswald (1994), gr-qc/9403036.
\item{$^{25}$} I. G. Avramidi, {\it New algebraic methods for calculating the
heat kernel and the effective action in quantum gravity and gauge theories},
University of Greifswald (1994),
gr-qc/9408028; in {\it Heat Kernel Techniques and Quantum Gravity, Discourses
in Mathematics and Its Applications},  No.~4, edited by  S.~A. Fulling, Texas
A\&M University, (College Station, Texas, 1995), to appear.
\item{$^{26}$} I. G. Avramidi, Phys. Lett. B {\bf 336}, 171 (1994).
\item{$^{27}$} I. G. Avramidi, {\it A new algebraic approach for calculating
the heat kernel in quantum gravity}, University of Greifswald (1994),
hep-th/9406047, submitted to J. Math. Phys.
\item{$^{28}$} I. G. Avramidi, {\it Covariant algebraic calculation of the
one-loop effective potential in non-Abelian gauge theory and a new approach to
stability problem}, University of Greifswald (1994), gr-qc/9403035, J. Math.
Phys. (1995), to appear.
\item{$^{29}$} A. Erdelyi, W. Magnus, F. Oberhettinger and F. G. Tricomi, {\it
Higher Transcendental Functions}, vol. I, (McGraw-Hill, New York, 1953).

\bye